\documentclass[%
twocolumn,
superscriptaddress,
 amsmath,amssymb,aps, %prl,
 floatfix, showpacs,
prb,
citeautoscript
]{revtex4-1}

\usepackage{graphicx}% Include figure files
\usepackage{dcolumn}% Align table columns on decimal point
\usepackage{bm}% bold math
\usepackage{amsfonts,amssymb,amsmath}
\usepackage{color}
\begin{document}

\preprint{}

\title{Non-colinear exchange interaction in transition metal dichalcogenide edges}
%\thanks{A footnote to the article title}%

\author{Oscar \'Avalos-Ovando}
 \email{oa237913@ohio.edu}
 \affiliation{Department of Physics and Astronomy, and Nanoscale and Quantum
 	Phenomena Institute, \\ Ohio University, Athens, Ohio 45701--2979, USA}
\author{Diego Mastrogiuseppe}
\affiliation{Department of Physics and Astronomy, and Nanoscale and Quantum
	Phenomena Institute, \\ Ohio University, Athens, Ohio 45701--2979, USA}
\affiliation{Instituto de F\'isica Rosario (CONICET), 2000 Rosario, Argentina}
\author{Sergio E. Ulloa}
\affiliation{Department of Physics and Astronomy, and Nanoscale and Quantum
	Phenomena Institute, \\ Ohio University, Athens, Ohio 45701--2979, USA}
 %This line break forced with \textbackslash\textbackslash
%

\date{\today}% It is always \today, today, but any date may be explicitly specified

\begin{abstract}
We study the Ruderman-Kittel-Kasuya-Yosida effective exchange interaction between magnetic impurities
embedded on the edges of transition-metal dichalcogenide flakes, using a three-orbital tight-binding model.
Electronic states lying midgap of the bulk structure have strong one-dimensional (1D) character, localized on the edges of the crystallite.
This results in exchange interactions with $1/r$ (or slower) decay with distance $r$, similar to other 1D systems.
Most interestingly, however, the strong spin-orbit interaction in these materials results in sizable non-collinear Dzyaloshinskii-Moriya interactions between impurities, comparable in size to the usual Ising and in-plane components.
Varying the relevant Fermi energy by doping or gating may allow one to modulate the effective interactions, controlling the possible helical ground state configurations of multiple impurities.
\end{abstract}

 %75.30.Hx Magnetic impurity interactions, 75.75.-c Magnetic properties of nanostructures, 75.70.Tj Spin-orbit effects

\maketitle

\emph{Introduction}. Manipulation of the electron spin is the cornerstone of potential spintronics devices.
Materials with strong spin-orbit coupling (SOC) arise as promising alternatives, as electrical or optical probes can be used to access the
spin degree of freedom, as
is the case in transition-metal dichalcogenides (TMDs) \cite{novoselov,geim,mak,xiao}.
The elemental unit in these materials is a stack of three atomic layers, where M atoms
(M = Mo, W) are sandwiched between two chalcogen X layers (X = S, Se), building MX$_{2}$ and providing a trigonal prismatic bonding environment to the metal atoms.

Exfoliation or direct CVD growth often produces nanoscale samples--nanoflakes--with different shapes and boundaries. One of the usual
flake geometries is triangular, often with irregularly shaped edges \cite{Chiu,vanderZande}.
Finite size samples, such as MoS$_{2}$ nanoribbons,
are predicted to exhibit unusual magnetic properties \cite{li,botello,tongay}, probably linked to the presence of extended states on zigzag edges.
Intrinsic magnetism in MoS$_{2}$ may arise \cite{zhang,lauritsen}, perhaps associated with edge states as those seen in STM images of flakes \cite{bollinger}. Polarization discontinuity effects are also predicted to result in interesting charged metallic
1D states for zigzag edges in TMDs \cite{Noguera,Marzari}. Quantum spin Hall effect has also been predicted in TMD-based structures with distorted 1D zigzag-chains\cite{qian}.

Magnetic impurities can interact effectively through an exchange process mediated by the conduction electrons of a host, the Ruderman-Kittel-Kasuya-Yosida (RKKY) interaction \cite{ruderman,kasuya,yosida}.
This has been studied on finite 2D materials, such as graphene nanoflakes \cite{szalowski1,szalowski3}, and nanoribbons \cite{szalowski2,black,black2,akbari,duffy}, where impurities lie close or on zigzag and armchair edges.  On graphene, RKKY interactions with dominant 1D character have been identified for impurities near sample edges, \cite{duffy} and line defects \cite{jiang}. Magnetic impurities, such as Mn, Fe\cite{cong,lu}, Co\cite{lu} or Ti\cite{andriotis}, can be introduced by STM and/or associated with Mo or S vacancies.

For TMDs, a combination of strong SOC and its coupling to the valley degree of freedom can generate exchange interactions with features that do not appear in conventional systems.
The strong SOC induces sizable anisotropic terms in the effective exchange interaction between impurities \cite{parhizgar,hatami,mastrogiuseppe},
including Dzyaloshinskii-Moriya (DM) terms.  The latter would result in non-collinear configurations of magnetic moments embedded in the 2D crystals, akin to impurities in heavy metals, where helical arrangements are reported \cite{Wiesendanger}.
The interaction in 2D TMD crystals is also found to depend strongly on the direction of impurity separation with respect to the underlying lattice, highlighting the importance of  crystal symmetries in the effective exchange interaction that results.

The unique properties of TMDs and the appearance of edge states extended along nanoribbons and flake edges prompt the question of how magnetic impurities interact in such unusual environment.  This motivates the study of RKKY interactions between impurities in TMD nanoflakes, with special attention to the effect of the 1D edge state features, and the role of SOC on the interaction.  Studies in 1D and 2D electron gases have
predicted twisted arrangements between magnetic impurities in the presence of ad hoc Rashba and Dresselhaus SOC \cite{imamura,zhu,flensberg}.

In this work we use a three-orbital tight-binding model \cite{liu} to study the interaction between two magnetic impurities in zigzag-terminated MoS$_{2}$ nanoflakes with different Fermi level, as provided by doping or gating of the sample \cite{dolui,laskar,mishra,cheng}. For Fermi levels at midgap energies, associated with states located at the flake edges, we explore the role of
1D and strong SOC on the effective exchange interaction between impurities.  We analyze the role of impurity separation
and explore strength and anisotropy of the
interaction as doping density or gating change.  We find the expected $1/r$ decay behavior of the
RKKY interaction envelope with distance $r$ for low doping levels, although Fermi levels deeper in the gap (higher doping)
exhibit markedly slower decay. More interestingly, we find that the
DM interaction terms that appear in this system are similar in strength to the usual anisotropic Heisenberg interactions,
which would give rise to interesting 1D helical configurations for impurities arranged on the edges of these flakes.

%\section{Model and Approach}

\emph{Model}. We study triangular zigzag-terminated MoS$_{2}$ nanoflakes with magnetic impurities hybridized near the edges of the flake. For
energies close to the optical gap, 4\emph{d}-orbitals from the Mo atoms contribute the most to the band structure \cite{liu}, with only slight admixture from the $p$-orbitals in the chalcogen atoms.  As such,
the system can be modeled by a triangular lattice of Mo atoms with multiple $d$-orbitals.
A suitable model at low energies around the gap is a three-orbital tight-binding (3O-TB) model, representing the dominant
$d_{z^2}$, $d_{xy}$ and $d_{x^2-y^2}$ orbitals \cite{liu}.
The Hamiltonian is $H = H_{\text{3O-TB}} + H_{\text{I}}$, where $H_{\text{3O-TB}}$ describes the TMD lattice and $H_{\text{I}}$
the two magnetic impurities of interest.  We write $H_{\text{3O-TB}} = H_{\text{o}} + H_{\text{t}}$, with
\begin{equation}\label{lattice2}
  H_{\text{o}} = \sum_{ \textbf{l}}^{N_{sites}} \sum_{s=\uparrow,\downarrow} \sum_{\alpha,\alpha'} \varepsilon_{\alpha\alpha',s}d_{\alpha,\textbf{l},s}^{\dagger}d_{\alpha',\textbf{l},s},
\end{equation}
where $d_{\alpha,\textbf{l},s}$ ($d^{\dagger}_{\alpha,\textbf{l},s}$) annihilates (creates) a spin-$s$ electron in orbital $\alpha$ $\in\,\left\{d_{z^2},d_{xy},d_{x^2-y^2}\right\}$ on site $\textbf{l}=l_{1}\textbf{a}_{1}+l_{2}\textbf{a}_{2}$;
$\textbf{a}_{1}=a(1,0)$, $\textbf{a}_{2}=a(1/2,\sqrt{3}/2)$
are lattice vectors of the triangular lattice with lattice constant \emph{a}.

\begin{figure}
  \includegraphics[width=0.25\textwidth]{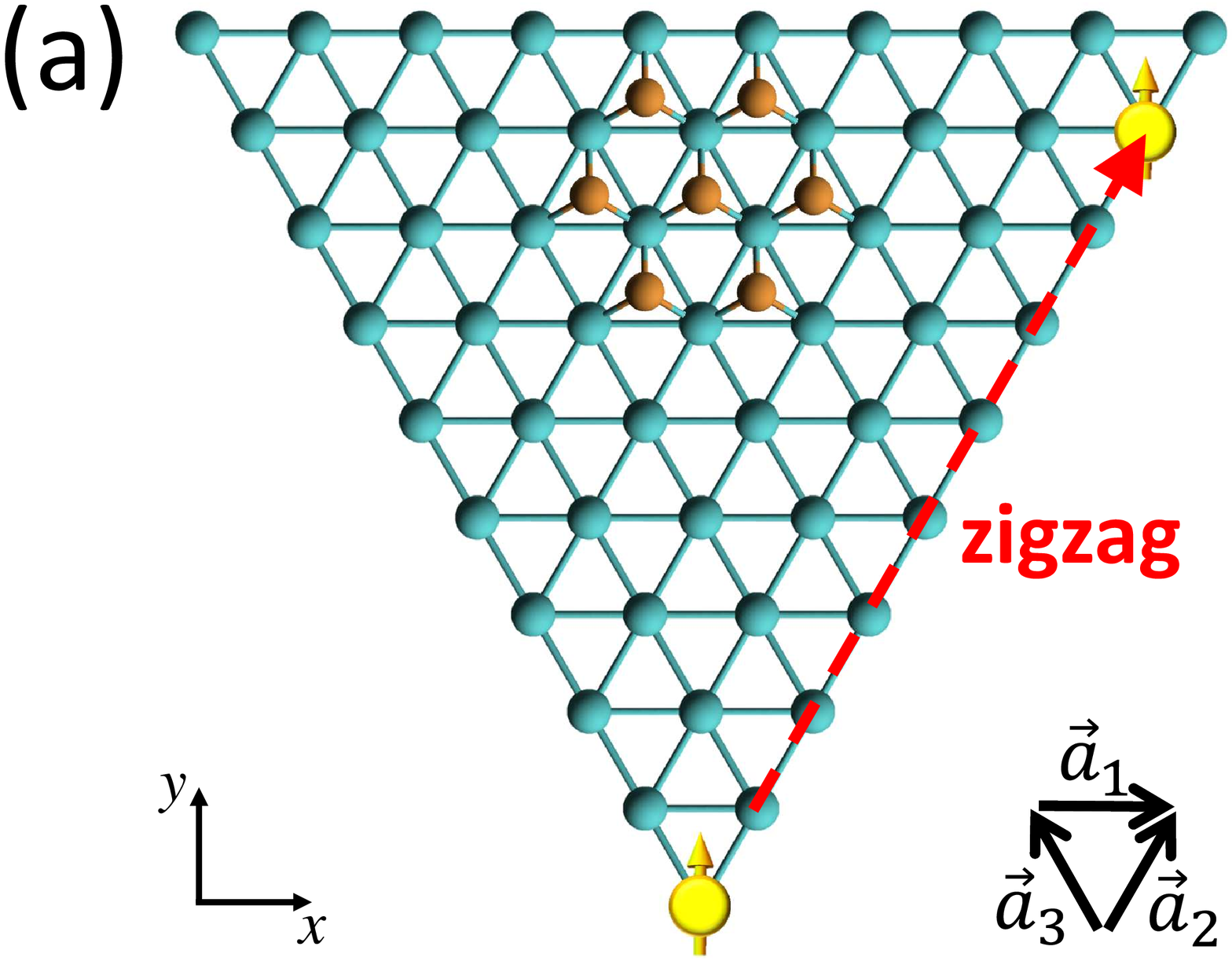}\\
  \includegraphics[width=0.47\textwidth]{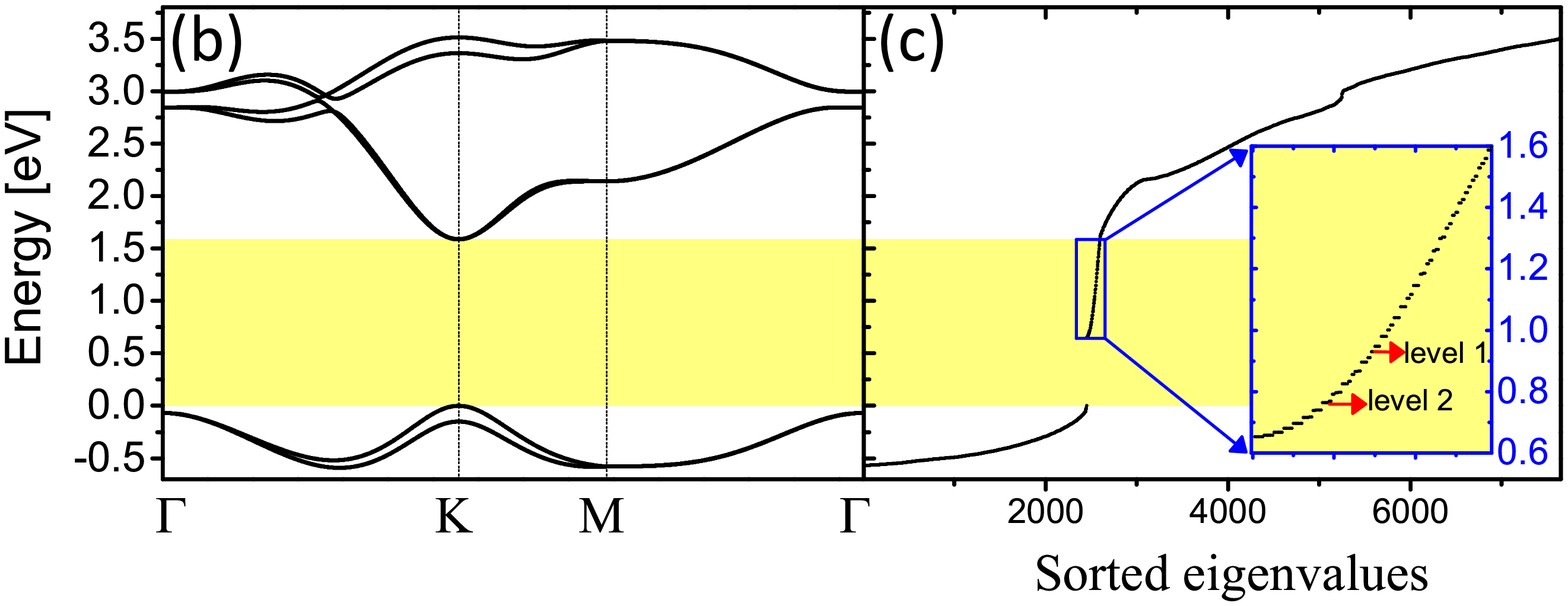}
  \caption{(Color online) (a) Top view of the triangular zigzag-terminated TMD nanoflake, where each site is a Mo atom
 (S atoms in dark orange shown only over small region for clarity). Two magnetic moments are
represented by yellow ball-arrows; here one is fixed at a corner, while the other is moved along the zigzag edge, red dashed line. (b) Band structure
of the MoS$_{2}$ 2D bulk monolayer, obtained from $H_{\text{3O-TB}}$; gap ($\sim1.6$ eV) highlighted in yellow.
(c) Discrete energy levels for 50-row flake, showing midgap edge states
present in the finite flake. Two different levels of doping (or gating) considered in this work are shown in inset.}\label{fig1}
\end{figure}

The nearest-neighbor hopping Hamiltonian is given by
\begin{equation}
H_{\text{t}} = \sum_{\textbf{l,a}_j} \sum_{s=\uparrow,\downarrow} \sum_{\alpha,\alpha'}
t_{\alpha\alpha'}^{(\textbf{a}_{j})}d_{\alpha,\textbf{l},s}^{\dagger}d_{\alpha',\textbf{l}+\textbf{a}_{j},s}+
\text{H.c.},\\
\end{equation}
where $t_{\alpha\alpha'}^{(\textbf{a}_{j})}$ are hopping parameters in the three different directions $j=1,2,3$, with $\textbf{a}_3=\textbf{a}_2 - \textbf{a}_1$.
The parameters in the Hamiltonian, including appropriate SOC terms, are taken from [\onlinecite{liu}] and [\onlinecite{pavlovic}],  where they are shown to provide
reliable description of the 2D bulk electronic band structure for energies close to the optical gap in several TMDs, see Fig.\ \ref{fig1}(b)\cite{footnote1}.
In a sample with edges, the level structure exhibits 1D-like extended states localized near the borders of the sample \cite{Carlos,Guinea},
and with energies in the gap region of the 2D crystal bulk.  In the finite-size triangular flakes we consider here, the electronic spectrum is
fully discrete, exhibiting both bulk- and edge-like states, the latter lying midgap and strongly
localized near the edges of the crystallite \cite{pavlovic,Carlos}.

The magnetic impurities in the system are assumed to hybridize with atomic \emph{d}-orbitals of the Mo atoms in the nanocrystal.  The
Hamiltonian for the magnetic impurities connected to specific sites of the TMD lattice is then
\begin{equation}\label{impurities1}
  H_{\text{I}}=\sum_{i=1,2} {\cal J}_\alpha \, \textbf{S}_{i}^{\text{imp}}\cdot\textbf{S}_{\alpha}(\textbf{l}_{i}),
\end{equation}
with local exchange coupling  ${\cal J}_\alpha$ between the impurity spin $\textbf{S}_{i}^{\text{imp}}$ and the electrons in orbital $\alpha$
at the location of the impurity, site $\textbf{l}_{i}$.
$\textbf{S}_{\alpha}(\textbf{l}_{i})=\frac{1}{2}\sum_{s, s'=\uparrow,\downarrow}d_{\alpha,\textbf{l}_{i},s}^{\dagger}%
\boldsymbol{\sigma}_{s,s'}d_{\alpha,\textbf{l}_{i},s'}$
is the  electron spin at site $\textbf{l}_{i}$ for orbital $\alpha$, where $\boldsymbol{\sigma}$ is the vector of spin-$\frac12$ Pauli matrices.
The electronic degrees of freedom are integrated out using second order perturbation theory, which yields the inter-impurity
effective exchange interaction
\begin{eqnarray}\label{jeffective1}
H_{RKKY} &=& J_{XX}\left(S_{1}^{x}S_{2}^{x}+S_{1}^{y}S_{2}^{y}\right)+J_{ZZ}S_{1}^{z}S_{2}^{z}\nonumber\\
 & &+J_{DM}\left(\textbf{S}_{1}\times \textbf{S}_{2}\right)_{z},
\end{eqnarray}
where  $J_{XX} = J_{YY}$ (in-plane), $J_{DM}$ (in-plane Dzyaloshinskii-Moriya) and $J_{ZZ}$ (Ising) terms are proportional to the static spin susceptibility tensor of the electron system \cite{ruderman,kasuya,yosida}. Behavior and characteristics of the different $J$'s in the 2D bulk have been discussed recently \cite{parhizgar, mastrogiuseppe}, finding them to be strongly dependent on the doping level and direction along which the impurities are located with respect to the crystal axes, among other features.

For finite flakes, the effective interaction is obtained from direct calculation of the difference between triplet and singlet
impurity configurations in the system ground state as $J_{\mathrm{eff}}/2=E(\uparrow\uparrow)-E(\uparrow\downarrow)$ \cite{deaven,black}.
The energy of the overall system, including magnetic impurities, is given by the sum of the sorted energy states of the full Hamiltonian
up to a given Fermi energy $\epsilon_{\text{F}}$,
$E(\textbf{S}^{\text{imp}}_{1},\textbf{S}^{\text{imp}}_{2})=\sum_{s=\uparrow,\downarrow}\sum_{i=1}^{\epsilon_{\text{F}}}\epsilon_{i,s}$,
as obtained by numerical diagonalization.

%\section{Results}

\emph{Results}. We consider \emph{p}-doped triangular flakes with 50 rows of metal atoms, corresponding to 1275 sites ($\simeq 160 \text{\AA}$ on
edge). We note that \emph{p}-doping has been achieved experimentally by substituting Mo by Nb in CVD-grown MoS$_{2}$ films \cite{dolui}, consistent with first principles studies\cite{laskar}. Other suggestions include replacing Mo by Mn\cite{mishra}, and doping with a wide class of transition metal atoms\cite{cheng}. We specify here characteristic doping levels inside the 2D energy gap [levels 1 and 2, $\varepsilon_{F1}> \varepsilon_{F2}$, as shown in the inset of Fig.\ \ref{fig1}(b)]. We use $\varepsilon_{F1}=0.9305$ eV and $\varepsilon_{F2}=0.7688$ eV,  corresponding to 26 and 52 holes in the nanocrystal flake, or $2.4\times10^{13}$ and $4.7\times10^{13}$ holes/cm$^{2}$, respectively. The bare couplings between localized and itinerant magnetic moments are set to ${\cal J}=0.3\text{ eV}$.

\begin{figure}
   \includegraphics[width=0.51\textwidth]{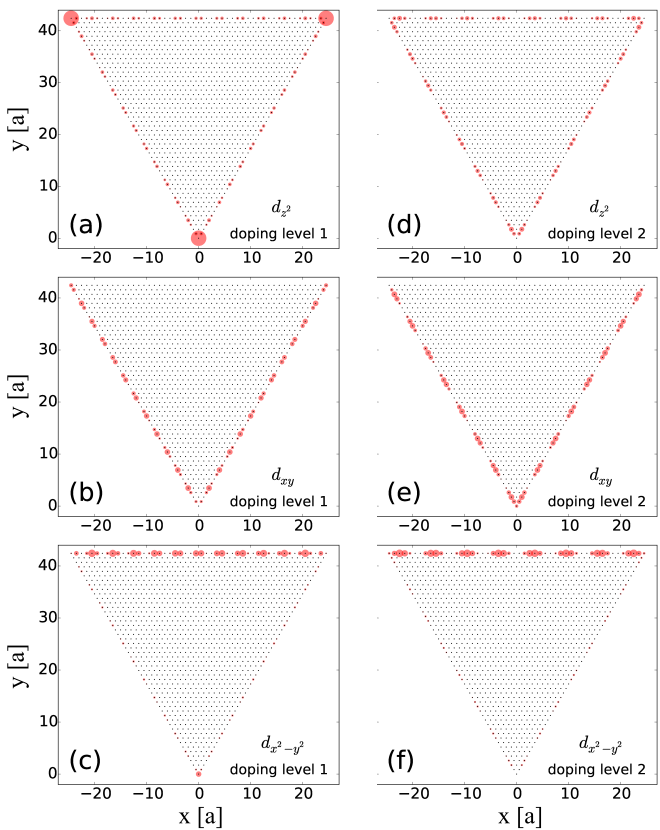}
  \caption{(Color online) Square magnitude of wave function components, as indicated by radius of red circles, for two in-gap levels: (a)-(c) level
1 in Fig.\ \ref{fig1} ($\varepsilon_{F1}=0.9305$ eV) and orbitals (a) $d_{z^2}$, (b) $d_{xy}$ and (c) $d_{x^2-y^2}$; (d)-(f) level 2
($\varepsilon_{F2}=0.7688$ eV) and orbitals as shown.
Black dots indicate Mo sites.
The $d_{xy}$ and $d_{x^2-y^2}$ components have been amplified by a factor of 3 with respect to those of the $d_{z^2}$ orbital.
}\label{fig2}
\end{figure}

Before exploring the resulting exchange interactions, we study the characteristics of the relevant eigenstates close to the Fermi level.
Figure \ref{fig2} shows maps of the square magnitude of the wave functions for the different orbitals and energies of interest.  These maps generally extend over the entire flake for energies well into the band continuum (not shown), with a
distribution not unlike that in extended 2D crystallites, although with slight spatial modulations due to finite size effects \cite{pavlovic,Carlos}.  However,
for levels in the midgap (such as levels 1 and 2 in Fig.\ \ref{fig1}(c) inset), the situation is very different.  There,
wave functions are predominantly along the edges, with a larger magnitude for the $d_{z^2}$ component than for the other
two orbitals (by a typical factor of 3). The midgap states are characteristically into groups of six nearly-degenerate states,
associated with the spin and three-fold symmetry of the flake. The wave functions in these groups have similar spatial patterns and show
strong localization on/near the border of the flake, while being extended along the edge.  Notice also an
oscillation pattern along the edge, as expected from the finite length of the flake, with increasing wave number for increasing energy.\cite{footnote2}
Some energy states exhibit highly localized states at the corners of the flake, as seen in Fig.\ \ref{fig2}(a) for level 1. The effective 1D
character of states in the gap is evident, not unlike states in carbon nanotubes \cite{Shenoy, Costa, Klinovaja} or graphene edges \cite{szalowski1,black,akbari,duffy}. However, as these states can be seen to arise from the mixing of 2D-bulk states with strong SOC, different states in the vicinity of a given
energy carry information on spin and spatial structure that result in subtle effective interactions between the embedded magnetic impurities.

\begin{figure}
  \includegraphics[width=0.46\textwidth,angle=0]{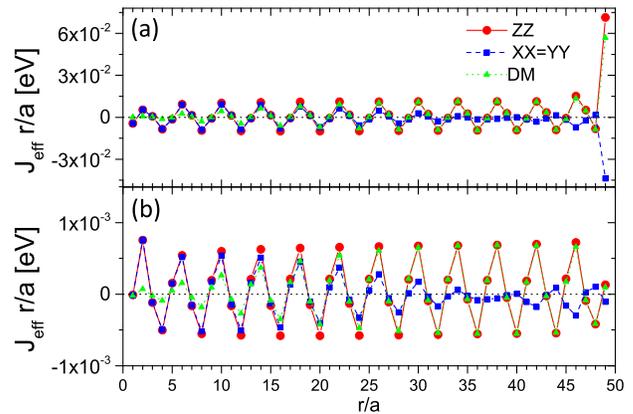}
  \caption{(Color online) Effective interactions [scaled by $(r/a)$] vs distance between impurities $r$, for level 1.
Red circles show the Ising interaction $J_{ZZ}$, blue squares the in-plane interaction $J_{XX}=J_{YY}$, and green triangles the DM interaction $J_{DM}$.
The first magnetic impurity is held at the lower corner of the flake while the second is moved along the zigzag edge, as shown in Fig.\ \ref{fig1}(a). First and second impurities are assumed hybridized (a) both to $d_{z^2}$ orbitals, and (b) to $d_{x^2-y^2}$ and $d_{xy}$ orbitals, respectively. The near
constant amplitude of the oscillations indicates an overall $1/r$ envelope. }\label{fig3}
\end{figure}

Figure \ref{fig3}(a) shows the effective interaction vs impurity separation for level 1, assuming both impurities hybridize to $d_{z^2}$
orbitals, in the configuration shown in Fig.\ \ref{fig1}(a). The interaction oscillates between ferromagnetic FM ($J<0$) and antiferromagnetic AFM ($J>0$) alignments, as expected,
with a period of $\simeq 4$ sites.
For shorter distances, up to the middle of the flake, the in-plane $J_{XX}$ and Ising $J_{ZZ}$ components are in phase and have similar magnitudes,
although the in-plane component decays further on.  Notice the presence of an additional spatial envelope of period $\simeq 38a$ for the in-plane interaction, as $J_{XX}$ starts mirroring the behavior for $r/a < 38$.
From the middle of the flake onwards, the DM and Ising components are in phase and have similar magnitudes. The envelope of
$J_{DM}$ is slightly out of phase with $J_{XX}$.  Notice that Fig.\ \ref{fig3} shows the exchange interaction scaled by
the spatial separation, so that a nearly constant amplitude of oscillation over the range shown illustrates the overall $1/r$ dependence of the interaction,
as one would expect for a purely 1D system \cite{Yafet, litvinov}.

The interaction is enhanced when both impurities are at the corners of the flake, associated with the large amplitude of the $d_{z^2}$ component
at the corners, Fig.\ \ref{fig2}(a). This suggests that
locating magnetic impurities near defects with large amplitudes of the local density of states at the Fermi energy would naturally enhance the interactions,
although of course such enhancement does not occur whenever the impurities are away from the corner defects.

Figure\ \ref{fig3}(b) shows the interactions when the first and second impurities hybridize to different orbitals, here $d_{x^2-y^2}$ and  $d_{xy}$, respectively. The orbital components at this level are also distributed along the edges, as shown in Fig.\ \ref{fig2}(b,c).
The interactions exhibit similar behavior to that described above, but now they are nearly one order of magnitude smaller, as the $d_{z^2}$
component is generally larger than for the other two orbitals.
One can also see a similar envelope function with $\simeq 1/r$ dependence modulating a 4-site periodic pattern, with a DM component
comparable in size to the other two components over the entire range shown.

\begin{figure}
  \includegraphics[width=0.46\textwidth,angle=0]{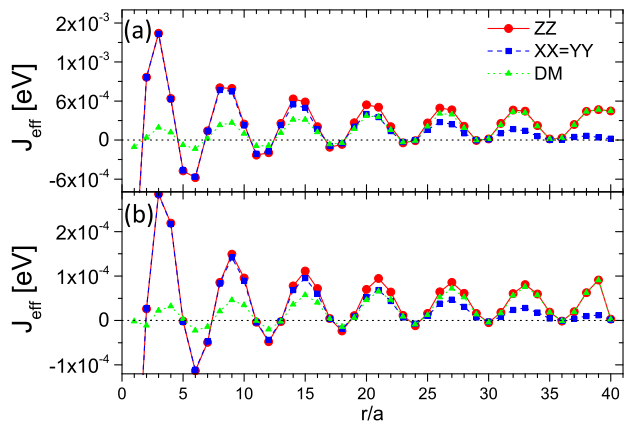}
  \caption{(Color online) Effective interaction vs relative impurity separation $r$, for level 2. Red circles show Ising interaction $J_{ZZ}$,
blue squares the in-plane interaction $J_{XX}=J_{YY}$, and green triangles the DM interaction $J_{DM}$. The first magnetic impurity is held on
the edge, 10 rows from the corner of the flake, while the second is moved along the zigzag edge. First and second impurities are hybridized
(a) both to $d_{z^2}$ orbitals, and (b) to $d_{z^2}$ and $d_{x^2-y^2}$ orbitals, respectively.  The oscillations decay with an approximate
$1/\sqrt{r}$ dependence, much slower than for $\varepsilon_{F1}$ in Fig.\ \ref{fig3}.}
\label{fig4} \end{figure}

A notable and characteristic feature of the effective interaction for impurities embedded in this material is that the $J_{DM}$ component is
sizable.  This behavior, seen also in bulk 2D crystals, can
be traced back to the large SOC in TMDs \cite{mastrogiuseppe}.  The DM interaction, absent (or negligible) in graphene or carbon
nanotubes, appears here as having the same order of magnitude as $J_{ZZ}$ and $J_{XX}$.

Let us now examine the effective exchange interaction for a different energy (doping/gating) level in the midgap region, such as level 2 in
Fig.\ \ref{fig1}(c). Here, the first impurity sits on the edge, ten lattice constants away from the corner, while the second impurity is displaced along the
edge. When both impurities are connected to  $d_{z^2}$ orbitals,
the interaction shows a larger period of oscillation, now $\simeq 7$ sites, Fig.\ \ref{fig4}(a).  Unlike the case in Fig.\ \ref{fig3},
$J_{ZZ}$ and $J_{DM}$ decay somewhat much more slowly than $1/r$ (nearly as $1/\sqrt{r}$, as obtained from fits).
Figure \ref{fig4}(b) shows the interaction when the first impurity is connected to $d_{z^2}$ and the second to $d_{x^2-y^2}$ orbitals, respectively.
This cross-hybridization interaction shows similar slower decay, but the overall magnitude is smaller due to the weaker amplitude for the
$d_{x^2-y^2}$ orbital on the edge of the flake, Fig.\ \ref{fig2}(e,f).
The different terms exhibit similar behavior as before, sizable $J_{DM}$ that
becomes comparable to $J_{XX}$ and $J_{ZZ}$ for larger separations.

Let us discuss the spatial pattern seen for different doping levels. Although the modulation of the
different wave function components may play a role in strongly localized cases (such as the corner defects seen for level 1), the overall spatial pattern
of the different exchange terms is not determined by a single wave function.  Instead, as the impurities interact through electron scattering events with
all the states in the system, the overall decaying envelope and
oscillation frequency depend smoothly on the Fermi energy value. One notices, in particular, longer ranges as the Fermi energy moves
deeper into the 2D bulk gap (e.g. Fig.\ \ref{fig4}).
Moreover, when the impurities sit at locations of nearly vanishing wave function amplitude (such as away from the borders)
the interactions are substantially weaker, but the modulation pattern remains (not shown). All this confirms that the resulting spatial
pattern of the effective exchange subsumes the contributions of the entire state manifold, even as those near the Fermi level may
have a more important contribution, as anticipated in the continuum.

%\section{Conclusions}

\emph{Conclusions}. We have studied the indirect interaction between magnetic impurities hybridized to the edges of zigzag-terminated MoS$_{2}$ flakes,
for Fermi energies lying in the gap of the bulk 2D crystal.  These energies identify states that are strongly localized on/near the edges of the flake, and exhibit
similar characteristics to those observed via STM measurements in TMDs \cite{bollinger}. In general, the interactions show oscillations with a period of a few sites,
and decaying typically as $1/r$, although more slowly for deeper Fermi levels.  Most interactions are nearly in phase, except for
$J_{XX}$ that dephases from the rest at larger inter-impurity separations.
The alternation of AFM and FM alignment of the impurities is as expected and seen in other 1D systems.
However, here is that the Dzyaloshinskii-Moriya interaction is large and generally comparable to the other terms.
All these results predict the RKKY interaction for impurities along TMDs edges to have rather long range, and to result in helical
ground state ordering of magnetic impurities.
These interactions can in principle be tailored by doping or gating of the material to a suitable midgap state, and by judicious placement
of the impurities, as possible by STM manipulation, for example. \cite{khajetoorians}
Although these controls are experimentally challenging, they would provide unique control to enhance the interactions of impurities, which could be
further explored by local probes, such as spin-polarized STM \cite{Wiesendanger,khajetoorians}.

%\begin{acknowledgments}
\emph{Acknowledgments}. We acknowledge support from NSF grant DMR 1508325 and useful discussions with M. M. Asmar.
%\end{acknowledgments}


\begin{thebibliography}{99}
\bibitem{footnote2} For instance, level 2 states show $\simeq 8$ antinodes, while level 1 states have 12 antinodes along the edge, see Fig.\ \ref{fig2}.
\end{thebibliography}
\end{document}